\documentclass[prd,twocolumn,showpacs,amsfonts,amssymb]{revtex4}

\usepackage{graphicx}

\usepackage{bm}

\newcommand{\beq}{\begin{equation}}
\newcommand{\eeq}{\end{equation}}
\newcommand{\beqs}{\begin{eqnarray}}
\newcommand{\eeqs}{\end{eqnarray}}

\newcommand{\gsim}{\mathrel{\raisebox{-
.6ex}{$\stackrel{\textstyle>}{\sim}$}}}

\newcommand{\drawsquare}[2]{\hbox{%
\rule{#2pt}{#1pt}\hskip-#2pt
\rule{#1pt}{#2pt}\hskip-#1pt
\rule[#1pt]{#1pt}{#2pt}}\rule[#1pt]{#2pt}{#2pt}\hskip-#2pt
\rule{#2pt}{#1pt}}
\newcommand{\fund}{\raisebox{-.5pt}{\drawsquare{6.5}{0.4}}}
\newcommand{\asym}{\raisebox{-3.5pt}{\drawsquare{6.5}{0.4}}\hskip-6.9pt%
        \raisebox{3pt}{\drawsquare{6.5}{0.4}}}

\begin{document}

\title{Determination of SU(4)$_{TC}$ Technicolor Gauge Group from Embedding 
in Extended Technicolor} 

\author{Masafumi Kurachi$^a$, Robert Shrock$^b$, and Koichi Yamawaki$^a$}

\affiliation{(a) Kobayashi-Maskawa Institute for the Origin of Particles and
  the Universe (KMI), Nagoya University, Nagoya 464-8602, Japan}

\affiliation{(b) C. N. Yang Institute for Theoretical Physics \\
Stony Brook University, Stony Brook, NY 11794 }

\begin{abstract}

In technicolor theories using an SU($N_{TC}$) gauge group, the value of
$N_{TC}$ is not, {\it a priori}, determined and is typically chosen by
phenomenological criteria. Here we present a novel way to determine $N_{TC}$
from the embedding of a one-family technicolor model, with fermions in the
fundamental represention of SU($N_{TC}$), in an extended technicolor theory,
and use it to deduce that $N_{TC}=4$ in this framework.  

\end{abstract}

\pacs{11.15.-q,12.60-i,12.60.Nz}

\maketitle

The possibility that electroweak symmetry breaking (EWSB) may occur
dynamically, as in technicolor (TC) theories \cite{tc} remains, although the
Higgs-like scalar boson discovered at the Large Hadron Collider (LHC) with mass
$m_H \simeq 125$ GeV \cite{lhc_atlas,lhc_cms} is consistent with being the
Standard Model (SM) Higgs.  A TC theory features an asymptotically free,
vectorial TC gauge symmetry and a set of massless TC-nonsinglet, SM-nonsinglet
fermions, $\{ F \}$. The TC theory becomes strongly coupled at the TeV scale,
confining and producing technifermion condensates $\langle \bar F F \rangle$,
with associated spontaneous chiral symmetry breaking (S$\chi$SB) and dynamical
EWSB.  Three of the resultant Nambu-Goldstone bosons (NGBs) are absorbed to
give masses to the $W^\pm$ and $Z$. To give masses to SM fermions, one embeds
the TC theory in a larger, extended technicolor (ETC) theory \cite{etc}.
Reasonably ultraviolet (UV)-complete ETC theories have been constructed as
asymptotically free chiral gauge theories that self-break \cite{sb} in stages
down to the (vectorial) TC subsector, yielding the generational hierarchy of SM
fermion masses, including light neutrino masses \cite{at}-\cite{ckm}.

Viable TC theories exhibit a squared gauge coupling $\alpha_{TC}(\mu) =
g_{TC}(\mu)^2/(4\pi)$ that grows to $O(1)$, but runs very slowly (walks) over a
substantial interval of Euclidean momenta $\mu$ and an associated large
anomalous dimension $\gamma_m \sim O(1)$ for the bilinear technifermion
operator \cite{wtctd}-\cite{wtc2}. These properties of a walking TC (WTC)
theory follow naturally if the theory has an approximate infrared (IR) zero in
the TC beta function $\beta_{TC}$ at a value $\alpha_{IR}$ that is slightly
larger than the critical minimal value, $\alpha_{cr,\bar F F}$, for the
formation of the $\langle \bar F F \rangle$ condensates \cite{b2}-\cite{irfp}.
Since $\alpha_{cr,\bar F F} \sim O(1)$, it is useful to calculate $\beta_{TC}$
and $\alpha_{IR}$ to higher-loop order \cite{bvh}.  Indeed, lattice studies
have shown that walking behavior can occur nonperturbatively even if the
perturbative beta function does not exhibit an IR zero
\cite{npwalking,su3_nf8}.  These $\langle \bar F F \rangle$ condensates
spontaneously break the approximate scale invariance of the TC theory, giving
rise to a light pseudo-NGB (PNGB), the technidilaton (TD), $\phi$
\cite{wtctd,td2,td_other}.  Using holographic methods, it has been shown that
WTC theories may yield a light TD \cite{td_hol1,td_hol2}. These holographic
studies extend earlier analyses using Schwinger-Dyson and Bethe-Salpeter
equations \cite{td_sdbs}.  Recent lattice studies of (vectorial) SU(3) gauge
theories with $N_f=8$ Dirac fermions (which is the value of $N_f$ in the
one-family TC (1FTC) model discussed here) have observed walking behavior and a
light composite TD-like scalar \cite{npwalking,td_lat_su3_nf8} (see also
\cite{td_lat_su3_nf12,latreview}).  The technidilaton in a WTC theory appears
to be consistent, to within theoretical and experimental uncertainties, with
currently measured properties of the Higgs-like scalar discovered at the LHC
\cite{td_hol1,td_hol2}, \cite{my_lhc}-\cite{sekhar_pdg} (although these
properties are also consistent with the SM Higgs). Future data from the LHC
will constrain the TC/TD scenario for the Higgs-like scalar further.

In addition to the requirement that the composite TD-like scalar in technicolor
must be consistent with the observed Higgs-like scalar, TC/ETC theories are
also subject to a number of phenomenological constraints, including those from
precision electroweak data \cite{sparam}, limits on flavor-changing neutral
current (FCNC) processes, etc. Both continuum (e.g.,
\cite{td_hol1,scalc,decon}) and lattice studies
\cite{su3_nf8},\cite{td_lat_su3_nf8}-\cite{latreview} have shown that TC
corrections to the $W$ and $Z$ propagators (in particular, the $S$ parameter
\cite{sparam}) can be substantially reduced in a theory with walking and,
moreover, via ETC effects \cite{decon}. Further, explicit calculations in a
reasonably UV-complete ETC theory showed that residual approximate generational
symmetries suppress FCNC processes \cite{ckm}.

The simplest embedding of the TC theory in ETC is obtained if one takes the
technifermions to comprise one SM family \cite{fs}, since in this case
the ETC gauge bosons are SM-singlets and $[G_{ETC},G_{SM}]=\emptyset$.
A common choice for the TC gauge group is SU($N_{TC}$).  Further,
the simplest models in this class of TC/ETC theories have technifermions
transforming according to the fundamental representation, $\fund$, of the
SU($N_{TC}$) TC gauge group, since in this case one just extends the TC gauge
indices on various fields to be ETC gauge indices (see Eq. (\ref{n_etc})).
Therefore, we shall consider here a 1FTC model with TC gauge group SU($N_{TC}$)
and technifermions in the fundamental representation.  (We do not consider
topcolor or higher-dimensional TC fermion representations.)
The value of $N_{TC}$ is typically determined by phenomenological criteria,
such as minimizing technifermion loop corrections to the $W$ and $Z$
propagators or fitting the properties of the Higgs-like scalar. 

Here we determine $N_{TC}$ from the embedding of a 1FTC theory, with
technifermnions in the fundamental representation of SU($N_{TC}$), in a
specific ETC theory. We find $N_{TC}=4$ in this framework, which agrees with
the value preferred by the fit to the Higgs-like scalar in this type of theory
\cite{td_hol1,my_lhc}. In addition to its phenomenological application, this
provides a novel example of how the structure of a low-energy effective field
theory is determined by its ultraviolet completion \cite{caveat}.

The natural embedding of the TC theory in ETC is 
\beq
{\rm SU}(N_{TC}) \subset {\rm SU}(N_{ETC}) \ . 
\label{tc_embedding}
\eeq
The ETC theory gauges the SM generation index and combines it with the TC
gauge index in such a way that each generational multiplet of SM fermions of
a given type is combined with the 1FTC fermions with the same SM quantum
numbers into a fundamental representation of SU($N_{ETC}$). 
Hence, with $N_{gen.}$ denoting the number of SM fermion
generations, 
\beq
N_{ETC} = N_{gen.} + N_{TC} \ . 
\label{n_etc}
\eeq
This determines $N_{ETC}$ in terms of the known $N_{gen.}=3$ and a given value
of $N_{TC}$ based on the embedding (\ref{tc_embedding}).  This procedure was
used with the minimal value, $N_{TC}=2$, to construct SU(5) ETC theories in
\cite{at,nt,ckm}. In \cite{at,nt,ckm}, given the value $N_{TC}=2$, the 
fermion content of the ultraviolet completion was chosen to yield the known 
$N_{gen.}=3$ \cite{difgen}.

We proceed to show how the embedding (\ref{tc_embedding}) of a 1FTC model, with
technifermions in the fundamental representation of the TC gauge group, in a
specific ETC theory with Eq. (\ref{n_etc}) determines that $N_{TC}=4$.  The key
to our result is the specification of the fermion content of the ETC theory and
the use of the requirement that the ETC theory must be free of chiral gauge
anomalies. Interestingly, our result is independent of $N_{gen.}$ (where it is
implicitly understood that $N_{gen.}$ is small enough so that SU(3)$_c$ and
SU($N_{ETC}$) are asymptotically free.)  To make this manifest, we will keep
$N_{gen.}$ general in our derivation.

The gauge group at the ETC scale is ${\rm SU}(N_{ETC}) \otimes G_{SM}$, where
$G_{SM}={\rm SU}(3)_c \otimes {\rm SU}(2)_L \otimes {\rm U}(1)_Y$ is the SM
gauge group, with $Q_{em}=T_{3L}+(Y/2)$. The fermion content of the ETC theory
is composed of two sectors, namely SM-nonsinglets and SM-singlets.  The
SM-nonsinglet sector is determined by our 1FTC theory and the embedding
(\ref{tc_embedding}) with (\ref{n_etc}). The fields are indicated below in
Young tableaux notation, where $a$ is a color SU(3)$_c$ index,
$i=1,...,N_{ETC}$ is the ETC index, and the representation is
$R=(R_{ETC},R_c,R_I)_Y$, with $I$ and $Y$ the weak isospin and hypercharge:
\beq
Q_L^{ai} = \quad \left ( \begin{array}{c}
                           u^{ai} \\
                           d^{ai} \end{array} \right )_L \ : \
(\fund,\fund,\fund)_{1/3}
\label{ql}
\eeq
\beq
u^{ai}_R: \ (\fund,\fund,1)_{4/3} \ , \quad 
d^{ai}_R: \ (\fund,\fund,1)_{-2/3}
\label{qr}
\eeq
\beq
L_L^i = \quad \left ( \begin{array}{c}
                           \nu^i \\
                           \ell^i \end{array} \right )_L \ : \
(\fund,1,\fund)_{-1}
\label{ll}
\eeq
\beq
\nu^i_R: \ (\fund,1,1)_0 \ , \quad \ell^i_R: \ (\fund,1,1)_{-2} \ .
\label{lr}
\eeq
The indices $i=1,...,N_{gen.}$ are SM generation indices and 
$N_{gen.}+1 \le i \le N_{ETC}$ are TC indices; the $a=1,2,3$ are color
SU(3)$_c$ indices. Thus, for example, $u^{a1}
\equiv u^a$, $u^{a2} \equiv c^a$, $u^{a3} \equiv t^a$ for the charge 2/3
quarks, and so forth for the other SM fermions. 

The minimal set of SM-singlet, ETC-nonsinglet fermions (written as
right-handed) is as follows:
\beq
\psi^{ij}_R = \psi^{[ij]}_R: \ (\asym,1,1)_0
\label{psi}
\eeq
and
\beq
\chi_{i,s,R}: \ (\overline{\fund},1,1)_0 \ , \quad 1 \le s \le N_{ETC}-4 \ . 
\label{chi}
\eeq
where $s$ labels each of the $N_{ETC}-4$ copies (flavors) of $\chi_{i,s,R}$. We
denote the contribution of a chiral fermion of representation $R$ of SU($N$) to
the triangle anomaly as $Anom(R)$. Since
\beq
Anom(\asym)=(N-4) \, Anom(\fund) \ ,
\label{anomrel}
\eeq
our ETC theory is anomaly-free.  One can also add a vectorlike SM-singlet,
ETC-nonsinglet fermion subsector; here we restrict ourselves to the minimal
version of the model. 

For a given gauge group $G$, we denote the beta function
$\beta_{G}=d\alpha_G/d\ln \mu = -2\alpha_G \sum_{\ell=1}^\infty b^{(G)}_\ell
(\alpha_G/(4\pi))^\ell$.  The ETC theory is asymptotically free, with one-loop
beta function coefficient
\beq
b_1^{{\rm SU}(N_{ETC})} = \frac{1}{3}(9N_{ETC}-10) \ , 
\label{b1etc}
\eeq
so as the scale $\mu$ decreases from the UV toward the IR, the ETC squared
coupling $\alpha_{ETC}(\mu)$ grows. As $\mu$ decreases through a scale that we
denote $\Lambda_1$, $\alpha_{ETC}(\mu)$ increases through the minimal critical
value for the formation of a bilinear fermion condensate.  Using a vacuum
alignment argument \cite{vacalign}, we infer that this forms in the channel
\beq
\asym \times \overline{\fund} \to \fund  \ , 
\label{afbar_to_f}
\eeq
breaking SU($N_{ETC}$) to SU($N_{ETC}-1$). The associated condensate
is
\beq
\langle \sum_{j=2}^{N_{ETC}} \psi^{1j \ T}_R C \, \chi_{j,1,R} \rangle \ , 
\label{condensate1}
\eeq
where $C$ is the Dirac charge-conjugation matrix, and, by convention, we have
taken the ETC gauge index $i=1$ in $\psi^{ij}_R$ and the copy index $s=1$ in
$\chi_{j,s,R}$.  The fermions $\psi^{1j}_R$ and $\chi_{j,1,R}$ with $2 \le j
\le N_{ETC}$ involved in this condensate gain dynamical masses of order
$\Lambda_1$ and are integrated out of the low-energy effective theory (LEET)
applicable at scales $\mu < \Lambda_1$.  Of the fermions in Eqs. (\ref{psi})
and (\ref{chi}), the remaining nonsinglet ones in this SU($N_{ETC}-1$) LEET are
$\psi^{2j}_R$ with $3 \le j \le N_{ETC}$; and $\chi_{j,s,R}$ with $2 \le j \le
N_{ETC}$ and $2 \le s \le N_{ETC}-4$.  The $(2N_{ETC}-1)$ ETC gauge bosons in
the coset ${\rm SU}(N_{ETC})/{\rm SU}(N_{ETC}-1)$ gain masses of order $g_{ETC}
\Lambda_1 \simeq \Lambda_1$.  Diagrams involving exchanges of these massive ETC
vector bosons connecting SM fermions with technifermions produce masses for the
first generation ($i=1$) of SM fermions \cite{splits}.

This SU($N_{ETC}-1$) theory is again asymptotically free (with
$b_1^{{\rm SU}(N_{ETC}-1)}=(1/3)(9N_{ETC}-19)$)  so the
gauge coupling (inherited at $\Lambda_1$ from the SU($N_{ETC}$) theory) grows,
and we infer that at a somewhat lower scale, $\Lambda_2$, there is again
condensation in the channel (\ref{afbar_to_f}), breaking SU($N_{ETC}-1$) to
SU($N_{ETC}-2$).  The associated condensate is
\beq
\langle \sum_{j=3}^{N_{ETC}} \psi^{2j \ T}_R C \chi_{j,2,R} \rangle, 
\label{condensate2}
\eeq
where, by convention, we have taken the gauge index $i=2$ in $\psi^{ij}_R$ and
the copy index $s=2$ in $\chi_{j,s,R}$. The fermions $\psi^{2j}_R$ and
$\chi_{j,2,R}$ with $3 \le j \le N_{ETC}$ involved in this condensate gain
dynamical masses of order $\Lambda_2$ and are integrated out of the
SU($N_{ETC}-2$) LEET operative at $\mu < \Lambda_2$.  Of the fermions in
Eqs. (\ref{psi}) and (\ref{chi}), the remaining ones that are nonsinglets in
the SU($N_{ETC}-2$) LEET are $\psi^{3j}_R$ with $4 \le j \le N_{ETC}$; and
$\chi_{j,s,R}$ with $3 \le j \le N_{ETC}$ and $3 \le s \le N_{ETC}-4$.  The
$(2N_{ETC}-3)$ ETC gauge bosons in the coset ${\rm SU}(N_{ETC}-1)/{\rm
SU}(N_{ETC}-2)$ gain masses of order $\Lambda_2$.  Diagrams involving exchanges
of these massive vector bosons connecting SM fermions with technifermions
produce masses for the second generation of SM fermions \cite{splits}.

This sequential self-breaking of the SU($N_{ETC}$) theory continues
iteratively in $N_{ETC}-4$ stages, using the $N_{ETC}-4$ copies of 
$\chi_{j,s,R}$ fermions, so that the original SU($N_{ETC}$) (chiral) 
gauge symmetry is finally reduced to the (vectorial) SU($N_{TC}$) subgroup,
with the indices corresponding to the broken ETC symmetries being the SM
generation indices. Hence, 
\beq
N_{gen.} = N_{ETC}-4 \ . 
\label{ngenrel}
\eeq
Substituting this expression for $N_{gen.}$ into Eq. 
(\ref{n_etc}), we obtain the result
\beq
N_{TC}=4 \ .
\label{ntc4}
\eeq
This is our main result. We have determined $N_{TC}$ from the structure of the
specific ETC theory in which our 1FTC theory is embedded.  A particularly
intriguing aspect of our result is that, although Eq. (\ref{n_etc}) connects
$N_{ETC}$ and $N_{TC}$, it does so in a manner that involves $N_{gen.}$, but
our result is actually independent of $N_{gen.}$, provided that $N_{gen.}$ is
sufficiently small that the ETC theory is asymptotically free and breaks in the
indicated manner, and also that SU(3)$_c$ is asymptotically free.  Very
interestingly, our resulting value $N_{TC}=4$ agrees with the value inferred
from a fit to the properties of the Higgs-like scalar in a 1FTC theory
\cite{td_hol1,my_lhc,tcg,caveat}.

Henceforth, we set $N_{gen.}$ equal to the known value, $N_{gen.}=3$. Combining
this with Eqs. (\ref{n_etc}) and (\ref{ntc4}), we infer that the the ETC gauge
group is SU(7)$_{ETC}$.  As discussed above, this breaks in three stages to the
(vectorial) SU(4)$_{TC}$ group: ${\rm SU}(7)_{ETC} \to {\rm SU}(6)_{ETC} \to
{\rm SU}(5)_{ETC} \to {\rm SU}(4)_{TC}$.  The theory naturally accounts for the
mass hierarchy in the SM fermion generations, since the SM fermion masses in
the $i$'th generation result from exchange of ETC vector bosons with mass
$\Lambda_i$ and, in the ETC boson propagators, $\Lambda_1^{-2} \ll
\Lambda_2^{-2} \ll \Lambda_3^{-2}$.

The fermion content in the SU(4)$_{TC}$ theory consists of the SM-nonsinglet
fermions in Eqs. (\ref{ql})-(\ref{lr}) and the SM-singlet fermions in
Eqs. (\ref{psi})- (\ref{chi}) with $4 \le i, j \le 7$.  This TC beta is again
asymptotically free (with $b_1^{(TC)}=26/3$).  Hence, 
the TC coupling $\alpha_{TC}(\mu)$ inherited from the lowest ETC
theory, SU(5)$_{ETC}$ at $\Lambda_3$, continues to grow as $\mu$ decreases
below $\Lambda_3$ \cite{hc}. 

For a fermion condensation channel ($Ch$) $R_1 \times R_2 \to R_{Ch}$, a
measure of the attractiveness is $(\Delta C_2)_{Ch} = C_2(R_1) + C_2(R_2) -
C_2(R_{Ch})$, where $C_2(R)$ is the quadratic Casimir invariant.  A rough
estimate of $\alpha_{cr,Ch}$ is $\alpha_{cr,Ch} \simeq 2\pi/[3(\Delta
C_2)_{Ch}]$. The $\asym$ field is self-conjugate in SU(4)$_{TC}$ and, at a
scale $\Lambda_{AA}$ (where $A$ denotes the \underline{a}ntisymmetric rank-2
tensor, $\asym$) forms a condensate in the most attractive channel $\asym
\times \asym \to 1$, of the form $\langle \sum_{i,j,k,\ell=4}^7
\epsilon_{ijk\ell} \psi^{ij \ T}_R C \psi^{k\ell}_R \rangle$, with $(\Delta
C_2)_{AA} = 2C_2(\asym) = 5$. This is invariant under both SU(4)$_{TC}$ and
$G_{SM}$.  The next most attractive channel is $\fund \times \overline{\fund}
\to 1$ in TC, with $(\Delta C_2)_{\bar F F} = 15/4$, forming at the scale
$\Lambda_{\bar F F}$ and involving the condensates $\langle
\bar F F \rangle = \langle \bar F_L F_R \rangle + \ \langle \bar F_R F_L
\rangle$ with the $F$ technifermions in (\ref{ql})-(\ref{lr}).  The $\langle
\bar F F \rangle$ condensates produce EWSB. One has the rough estimate
\beq
\frac{\Lambda_{AA}}{\Lambda_{\bar F F}} \simeq 
\exp \bigg [ \frac{2\pi}{b^{(TC)}_1} (
 \alpha^{-1}_{cr,AA}- \alpha^{-1}_{cr,\bar F F} ) \bigg ] \ . 
\label{TCLambda_ratio}
\eeq
This yields $\Lambda_{AA}/\Lambda_{\bar F F} \simeq 1.5$. In general, a TD-like
scalar in this theory contains $\bar F F$, $AA$, a techni-glueball component,
etc.; the $\bar F F$ component plausibly dominates because of the
$\Lambda_{AA}/\Lambda_{\bar F F}$ ratio and the walking behavior, which implies
that the TD mass is $\ll m_{TC,had}$, where $m_{TC,had}$ denotes the mass 
scale of techni-vector mesons and techni-glueballs.

Neglecting ETC and SM gauge interactions, the 1FTC theory has a (non-anomalous)
global flavor symmetry involving SM-nonsinglet fermions, ${\rm SU}(8)_{F_L}
\otimes {\rm SU}(8)_{F_R} \otimes {\rm U}(1)_V$. This is broken to
${\rm SU}(8)_V \otimes {\rm U}(1)_V$ by the $\langle \bar F F \rangle$
condensate. In addition to the three NGBs absorbed by the $W^\pm$ and $Z$, this
yields 60 PNGBs. Taking account of walking and the strong ETC interactions, it
appears possible that their masses could be $\gsim$ O(1) TeV, above
current LHC limits \cite{mky_pngb}.

Although 1FTC with $N_{TC}=4$ has a large perturbative value of $S$, viz.,
$S_{pert.}=8/(3\pi)$, it is well-known that the perturbative estimate of $S$ is
not reliable because TC is strongly coupled at the scale of $m_W$ and
$m_Z$. Here, motivated by the results of \cite{td_hol1,scalc,decon}, we will
assume that the walking and ETC effects can suppress $S$ sufficiently to obey
experimental constraints.  Ongoing and future lattice calculations will further
test this assumption.  A constraint on TC/ETC models is that the spectrum of
technihadrons must be consistent with current limits from the LHC. We have
already commented on the PNGBs.  It also appears to be possible that the 1FTC
techni-vector meson masses may lie above the LHC limits of a few TeV
\cite{mky_tcvm}.  Additional ingredients are needed to fully explain the
spectrum of quark and lepton masses, in particular, $t$-$b$ mass splitting.

In summary, we have presented a novel way to determine $N_{TC}$ from the
embedding of a one-family SU($N_{TC})$ technicolor theory having technifermions
in the fundamental representation, in a particular SU($N_{ETC}$) extended
technicolor theory, with the SM fermions combined with technifermions into
fundamental representations of SU($N_{ETC}$) as specified in
Eqs. (\ref{ql})-(\ref{lr}) and have shown that this yields the value
$N_{TC}=4$.  This value is the same as one inferred from a fit to the 125 GeV
scalar boson in TC \cite{td_hol1,my_lhc}. Our result motivates lattice studies
of SU(4) gauge theory with $N_f=8$ Dirac fermions.  Future LHC data will yield
stringent tests of this model. In addition to this phenomenological
application, our result is of general interest for the insight that it provides
on how the structure of a low-energy effective field theory - here the TC
theory - is determined by its embedding in an ultraviolet completion, the ETC
theory.

This research was partly supported by the JSPS Grant-in-Aid for Scientific
Research (S) No. 22224003 and (C) No. 23540300 (K. Y.) and the U.S.  NSF Grant
No. NSF-PHY-13-16617 (R.S.). R.S. thanks the Kobayashi Maskawa Institute at
Nagoya University for warm hospitality during a visit when this research was
performed. 



\begin{thebibliography}{99}
\bibitem{npwalking}
%
While an IR zero in the perturbative beta function can yield walking behavior,
there are cases where the walking can occur nonperturbatively in the absence of
such an IR zero. This is shown by lattice studies of (vectorial) SU(3) theory
with $N_f=8$ Dirac fermions in the $\fund$ representation, which find walking
behavior \cite{su3_nf8}. For this theory, $b_1$ and $b_2$ have the same sign,
so the two-loop beta function has no IR zero (it has an IR zero for $9 \le N_f
\le 16$).
\bibitem{caveat}
%
Our results do not preclude other types of TC and ETC models, e.g., those with
different ETC/TC fermion content, additional strongly coupled gauge symmetries
or other composite-Higgs models. Future experimental data,
which at present is consistent with the minimal SM Higgs, will further
constrain all composite-Higgs models.
\bibitem{hc}
%
Although our minimal TC theory does not have a perturbative two-loop IR zero
in $\beta_{TC}$, one can add a vectorlike set of SM-singlet, (E)TC-nonsinglet
fermions producing such a zero and hence walking behavior from the two-loop
perturbative beta function (see also \cite{npwalking}).
\end{thebibliography}
\end{document}